\documentclass[aps,prb,twocolumn,groupedaddress]{revtex4}
\usepackage{graphics,epsfig}
\usepackage{dcolumn}

\begin{document}

\title{Evolution of Electronic and Vibrational Polarity of NaF Nanocrystals \\
from Diatomic to Bulk: A Density Functional Study}

\author{Christian Schmidt and Philip B. Allen }
\affiliation{Department of Physics and Astronomy, 
State University of New York, Stony Brook, NY 11794-3800}

\author{Tunna Baruah and Mark Pederson}
\affiliation{Center for Computational Materials Science, 
Naval Research Laboratory, Washington, DC 20375-5345}

\date{\today}

\begin{abstract}
Density functional theory (DFT) is used to study vibrations,
electrical dipole moments, and polarizabilities of
NaF clusters.  Because of prior experimental and theoretical
studies, this is a good model system for tracking the
evolution of the properties from diatomic molecule to bulk crystal.  
The ratio of vibrational to electronic contributions
to the polarizability increases dramatically with 
size $N$ in the closed shell clusters Na$_N$F$_N$.  The open shell
system Na$_{14}$F$_{13}$ has a greatly enhanced electronic
polarizability.   Contrary to previous studies on this system
which treated only the outer electron by quantum mechanics, 
we find the O$_h$ cubic structure
to be stable relative to the polar distorted structures such as C$_{3v}$. 
\end{abstract}


\maketitle

\section{\label{sec:l}Introduction}

Rayane {\it et al.} \cite{Rayane} have pioneered the
experimental study of dipole moments and polarizabilities
of NaF clusters, focussing on systems
such as Na$_{N}$F$_{N-1}$ with one electron outside a closed shell.  
However, even the closed shell systems Na$_{N}$F$_N$
have dielectric properties of considerable interest.
A study of this can illuminate such issues as how ferroelectric 
dipolar order of the bulk evolves up from
nanocrystals.  Our main focus is the quantity $\alpha_N$,
defined as the polarizability per molecular unit of a cluster of
$N$ molecular units,
\begin{equation}
\alpha_{N,\alpha\beta} \equiv \frac{1}{N}\partial \mu_{\alpha}/
\partial F_{\beta}
\label{eq:def}
\end{equation}
where $\vec{\mu}$ is the total dipole moment of a cluster or crystal
and $\vec{F}$ is the external electric field.

The Clausius-Mossotti relation
\begin{equation}
\alpha_{\infty}=\lim_{N\rightarrow\infty}\alpha_{N}
=\frac{\alpha_1}{1-(4\pi n/3)\alpha_1}
\label{eq:CM}
\end{equation}
is often used to relate the crystal polarizability per molecule $\alpha_{\infty}
=(\epsilon-1)/4\pi n$ to the molecular polarizability
$\alpha_1$ and the molecular density $n$. 
This equation is exact for a bulk crystalline
solid where {\bf point-polarizable} and isotropic molecules sit at sites of
cubic symmetry.  It takes into account the local electric
fields of all other induced dipoles which enhance the
applied field at the cubic sites.  
This provides the simplest model for a ferroelectric
instability when $\alpha_1$ is large enough.
Numerical linear algebra
allows the formula to be extended to less symmetric systems 
of point-polarizable molecules \cite{Allen}.  However,
no solid sequesters all polarizable entities accurately
into point objects, or has only dipolar interactions, so the
applicability of the formula to real ionic crystals,
while traditional, can be regarded as largely wrong \cite{Pantelides},
or alternately as basically right but only after the ions
are altered by crystal field and confinement effects \cite{Fowler}.
Here we use electronic structure theory (density functional
theory, or DFT) to provide ``data'' on the evolution of polarizability
\cite{Mahan}.

Semiconductors like Si \cite{Chelikowsky} and CdSe \cite{Efros}
have received a large amount of attention.  The alkali
halides, however, provide simpler model systems.  Hudgins
{\sl et al.} \cite{Hudgins}
have documented experimentally the stability of simple
cuboidal fragment structures in NaCl.  Much modeling has
been done, mostly using classical models such as the shell
model \cite{Welch1,Welch2,Phillips,Cheng,Doye,Calvo}.  
Electronic structure methods have been applied to 
diatomic alkali halide molecules \cite{Fowler2,Bacskay,Andrade} and
various alkali halide clusters \cite{Ahlrichs,Dickey,Modisette}.

We choose NaF as a model system for studying the 
evolution toward bulk properties as well as the
idiosyncrasies of individual nanoparticles.  
The bulk \cite{Buyers} and surface \cite{Brusdeylins}
vibrational properties are both known.
Foremost among idiosyncrasies of nanophases
is the fact that non-zero equilibrium electrical
dipoles $\vec{\mu}$ are necessary, just for reasons of cluster
geometry and symmetry.  This is well known \cite{Li,Krishnan} in materials
like CdSe whose bulk phase lacks inversion symmetry.
It is less well known that dipoles must occur
even in rather symmetrical cuboid
rocksalt fragments, unless they have high enough
symmetry (T$_d$ or symmetries which include inversion) 
to forbid a dipole.  

Concerning electrical polarity of nanoclusters,
there is a large literature of experimental and theoretical
studies for metal clusters, especially simple
metals \cite{deHeer,Bonin} ({\it e.g.} Na).  
There are also studies of polarity in 
semiconducting nanoclusters \cite{Jackson}.  
Microscopic dielectric screening theory has been developed and
implemented for semiconductors \cite{Ogut,Delerue}.  In ionic nanophase
materials no systematic study of polarity using full electronic
structure theory is known to us.  One new feature in ionic
materials, not seen in metals or homopolar semiconductors, is a large
contribution to polarity coming from nuclear displacements.
We call this contribution the ``vibrational'' part $\alpha_{\rm vib}$.
Heteronuclear materials with
partially ionic nature, such as the rocksalt crystal GeTe, can have
ferroelectric instabilities driven partly by vibrational
polarization \cite{Steigmeier}.

NaF is a good choice for a first
study for two reasons.  First, there is a lot of previous
work on structural properties of this and
related alkali halide clusters, for which structures are
generally simple fragments (nanocrystals, or ``cuboids'') of the bulk rocksalt
structure.  Second, because of the
small polarizability of the F$^{-}$ ion, and the
similar sizes of Na$^+$ and F$^{-}$ ions, NaF
should be simpler than all other alkali halides, potentially
more amenable to analysis using models in
the spirit of Clausius-Mossotti.
There have been surprisingly few calculations
using full electronic structure theory; 
non-stoichiometric varieties (which we largely ignore)
have received the most attention \cite{Bonacic}.
We postpone the study of classical pictures
for a future paper, and present here the results of a
quantum study using DFT.
This paper is based largely on the results of the MA
thesis of C. Schmidt \cite{Schmidt}.

\section{diatomic molecule}

\begin{figure}[top]
\includegraphics[angle=0,width=0.4\textwidth]{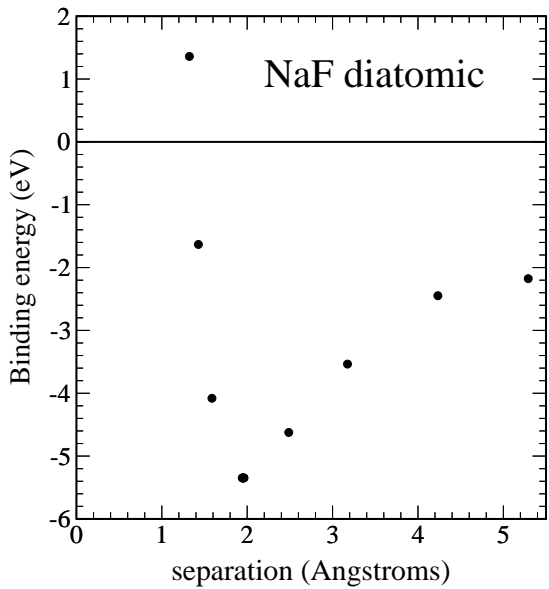}
\caption{\label{fig:gsen} Density functional total energy
of NaF diatomic molecule relative to free atoms.}
\end{figure}

The NaF diatomic molecule or ``monomer'' allows a
good comparison between our calculations and both
experiment \cite{Douay,Huber,Ekstrom} and previous theory 
\cite{Garcia,Bacskay,Dickey,Modisette}.
We use the package NRLMOL \cite{Pederson} which allows
versatile and accurate DFT calculations
for molecules.  We choose the PBE \cite{PBE} version of 
generalized gradient approximation (GGA),
and include a spin density functional for
systems with uncompensated spins ({\it i.e.} the Na and F
atoms and Na$_{14}$F$_{13}$.)  Results are shown in
Table \ref{tab:diatomic}.  We also show in Fig. \ref{fig:gsen}
the total energy from DFT relative to isolated atoms.
The experimental dissociation energy of 5.3 eV \cite{Ham}
is quite accurately reproduced, as are the bond-length,
dipole moment, and vibrational frequency.

\begin{table}
\caption{\label{tab:diatomic} Experimental and theoretical
properties of the NaF diatomic molecule.}
\begin{ruledtabular}
\begin{tabular}{lccccc}
&present calc.& expt. & expt. & theory & theory\\
&NRLMOL&ref.\onlinecite{Douay}&ref.\onlinecite{Huber}&
ref.\onlinecite{Dickey,Modisette,Andrade} 
&ref.\onlinecite{Fowler2,Bacskay}\\
\hline
$d \ (a_B)$& 3.69 & 3.64 & 3.64 & 3.6 & \\
$\mu \ (ea_B)$& 3.07 & 3.21 & &3.22 & 3.2 \\
$\omega$ (cm$^{-1}$)& 510 & 536 & 536 & 540 &\\
$\alpha_{{\rm el},\perp}$  (\AA$^3$)& 2.39 &&&1.82 &\\
$\alpha_{{\rm el},\parallel}$  (\AA$^3$)& 3.33 &&&1.95 &\\
$\alpha_{{\rm vib},\parallel}$  (\AA$^3$)& 0.86 &&&1.04 &\\
\end{tabular}
\end{ruledtabular}
\end{table}

\begin{table}
\caption{\label{tab:atomic} Experimental and theoretical
polarizabilities (in \AA$^3$) of Na and F atoms and ions.
SCF means self-consistent Hartree-Fock, and ACPF 
(average coupled pair functional) includes
correlation.}
\begin{ruledtabular}
\begin{tabular}{lcccc}
&Na& Na$^+$ & F & F$^-$ \\
\hline
NRLMOL& 23.18 & 0.144 & 0.536 & 1.23 \\
expt. ref \onlinecite{Ekstrom} & 24.11 &&&\\
SCF ref.\onlinecite{Bacskay}&& 0.139 && 1.58 \\
ACPF ref.\onlinecite{Bacskay}&& 0.148 && 2.84 \\
\end{tabular}
\end{ruledtabular}
\end{table}

Can the diatomic results be related to
atomic, ionic, or bulk properties?  Computed
atomic and ionic polarizabilities are shown in Table \ref{tab:atomic}.
The sum of DFT polarizabilities of the Na$^+$ and F$^-$ ions is
1.37 \AA$^3$.  This is close to the polarizability $\alpha_{\infty}=1.45$
per formula unit of the solid.  However, Clausius-Mossotti theory
gives the value $\alpha_1$=1.16 \AA$^3$ from the crystal data.
It is difficult to know
how to interpret these numbers.  Fowler and Madden \cite{Fowler}
make three relevant observations.  (1) Correlation corrections
greatly enhance the polarizability of the larger F$^-$ ion,
doing little for the isoelectronic but smaller Na$^+$ ion.
(2) Putting the ion in a crystalline environment makes the F$^-$
ion shrink, both because of the attractive crystal field of the
surrounding ions and because of the repulsion of the surrounding
electrons.  The resulting smaller ion has a polarizability which
is much less altered by correlation.  (3) The resulting smaller
polarizability estimated for the ``ion-in-a-crystal'' is perhaps
quite consistent with Clausius-Mossotti.
It is unfortunate that there are no direct experimental data for
the polarizatility of the ions Na$^+$ or F$^-$.  A long list
of highly diverse values for the F$^-$ polarizability
is tabulated by Iwadate and Fukushima
\cite{Iwadate}.  None are direct measurements.

It is difficult to know how to relate these ideas and numbers
to our calculated polarizability of the diatomic.
In zeroth approximation the equilibrium dipole of the diatomic is 
$ed$ and the electronic polarizability 
is the sum of the ionic polarizabilities, or 1.37\AA$^3$.
The computed dipole is smaller by 17\%, but the computed
polarizability is far larger
than the zeroth order estimate.
Using the average $\bar{\alpha}=(\alpha_{xx} +\alpha_{yy}+\alpha_{zz})/3$,
the diatomic has $\bar{\alpha}_{1,{\rm el}}$=2.70, larger by 2 than 
the sum of free ions.  Also the diatomic has anisotropy 
$\Delta\alpha/\bar{\alpha}$=34\%, where $\Delta\alpha^2=
[(\alpha_{xx}-\alpha_{yy})^2 + (\alpha_{yy}-\alpha_{zz})^2 + 
(\alpha_{zz}-\alpha_{xx})^2]/2$.  
In an attempt to account for these discrepancies,
three corrections were tested.  (1.) Each ion
polarizes in the self-consistent field of the other, and these local field
corrections are accounted by solving two coupled linear equations
for the moments $\mu_1$ and $\mu_2$.  Unlike Fowler and Madden \cite{Fowler},
we did not correct the polarizability of the ions for their
altered environments.  The corrections depend on the
ratios $\phi_{\pm}=\alpha_{\pm}/d^3$ for the two ionic species, 
where $d$ is the internuclear separation.  
For Na$^+$  $\phi_+$ = 0.0193, and  
for F$^-$ $\phi_-$ = 0.1652.  The reduced ratio
$\phi_{\rm red}=\phi_+ \phi_- /(\phi_+ + \phi_-)$ is 0.0173.
The corrected values of the permanent dipole \cite{Rittner}
and polarizability tensor \cite{Fowler2} are
\begin{eqnarray}
\mu &=& ed\left[\frac{1-\phi_+ -\phi_- -8\phi_+ \phi_-}
{1-4\phi_+ \phi_-}\right] 
=0.80ed \nonumber \\
\alpha_{\parallel} &=& (\alpha_+ + \alpha_-)
\left[\frac{1+4\phi_{\rm red}}{1-4\phi_+ \phi_-}\right] 
=1.08(\alpha_+ + \alpha_-) \nonumber \\
\alpha_{\perp} &=& (\alpha_+ + \alpha_-)
\left[\frac{1-2\phi_{\rm red}}{1-\phi_+ \phi_-}\right] 
=0.97(\alpha_+ + \alpha_-) \nonumber \\
\label{eq:diat}
\end{eqnarray}
The result of simple screening of the naive dipole $ed$ by point
ionic polarizability agrees closely with the computed total dipole
(Table \ref{tab:diatomic}), but
the enhancement of $\alpha$ parallel to the molecular axis, and
smaller enhancement perpendicular, 
while in the right direction, are much too small.  
It appears that when higher-order multipoles are included,
they make corrections comparable to the dipolar ones \cite{Klemperer}
so the agreement with the total dipole moment is partly accidental.
(2.) Nonlinear corrections are not negligible since the local
fields are large.  Since the ions are inversion-symmetric,
the leading correction is $\mu=(\alpha+\beta F^2 /2)F$.  Using
NRLMOL we computed $\beta$=7.9 and 0.06 for F$^-$ and Na$^+$,
in atomic units (a$_B^7$/e$^2$.)  The ions sit in fields
$F=e/r^2$ which are large, and the non-linear correction is
estimated to be +12\% for the polarizability of the F part of
NaF, and +1\% for the Na part; again the direction is right but
the effect is much too small.  (3.) There is incomplete charge
transfer between Na and F in the diatomic.  Using the computed
zero-field dipole as a measure, the static charge $\mu/d$ is $\pm 0.83 e$,
which suggests that the Na could be treated as 83\% Na$^+$
and 17\% Na.  Because of the giant polarizability of neutral Na 
(see Table \ref{tab:atomic}),
it is easy to rationalize a large correction, but less easy
to make a simple classical model for both tensor components of the
electronic polarizability.

\par
\par

The vibrational part of the polarizability is a new effect 
that occurs in the diatomic and has no counterpart in atoms or ions.
We compute it from the formula \cite{Vib,Baruah}
\begin{equation}
\alpha_{{\rm vib},\alpha\beta}=
\sum_{i\mu,j\nu} Z_{\alpha,i\mu}(K^{-1})_{i\mu,j\nu}Z^{\rm T}_{i\nu,\beta}
\label{eq:avib}
\end{equation}
where $Z_{\alpha,i\mu}$ is the dynamic effective charge tensor
$\partial\mu_{\alpha}/\partial u_{i\mu}$, $u_{i\mu}$ is the
$\mu$-th Cartesian component of the displacement of the $i$-th atom,
and $K$ is the force constant tensor.

Since atoms and ions have no vibrational polarizability,
we look instead to the bulk solid for guidance,
and here a surprise occurs.  The static $\epsilon_{\rm tot}=\epsilon_{\rm dc}$ 
and optical $\epsilon_{\rm el}=\epsilon_{\rm opt}$ dielectric constants of NaF
are 5.1 and 1.74 \cite{Tessman}; the effective polarizabilities
$\alpha=(\epsilon-1)/4\pi n$ are 8.04 \AA$^3$ (electronic plus
vibrational) and 1.45 \AA$^3$ (electronic only).  Thus the
vibrational contribution is 5 times the electronic
contribution to the bulk polarizability, whereas in the
diatomic molecule according to Table \ref{tab:diatomic}
the ratio is 0.29/2.70, 50 times smaller (this uses the
average of the diagonal elements of the $\alpha$ tensor.)
An alternate analysis of the bulk is to extract ``molecular''
polarizabilities from the Clausius-Mossotti formula.  These
are 1.16 \AA$^3$ (electronic) and 2.29 \AA$^3$ (vibrational).
The ratio is reduced from 5 to 2 but is still 20 times larger
than in the diatomic molecule.  This remarkable difference
comes partly because the electronic polarizability is reduced
(per molecule) in the bulk, but mostly because the vibrational
part is increased.

\section{clusters}

\begin{figure}
\includegraphics[angle=0,width=0.45\textwidth]{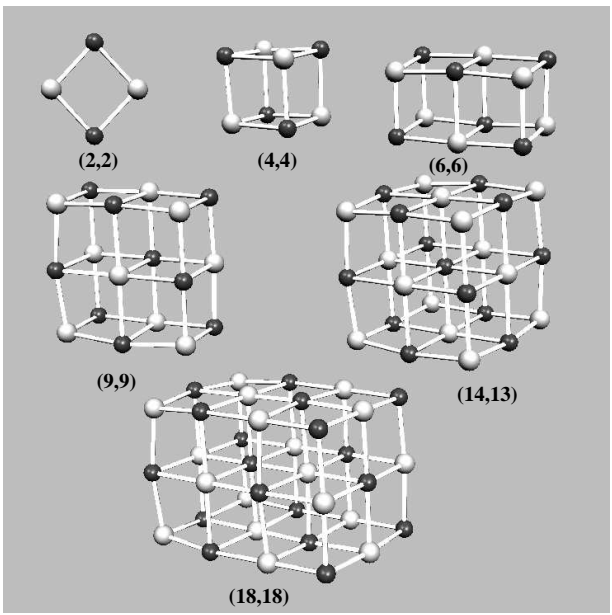}
\caption{\label{fig:struc}Structures of NaF rocksalt
fragments (cuboids, except for Na$_2$F$_2$) given by DFT.  
The Na$_4$F$_4$ ring structure
and O$_h$ $4\times 4\times 4$
cuboid Na$_{32}$F$_{32}$ are not shown.  The Na$_9$F$_9$ and
Na$_{18}$F$_{18}$ C$_{4v}$ structures have even numbers
of $3\times$3 planes; planes are charged alternately + or -,
so the fragments are polar.  The Na$_{14}$F$_{13}$ O$_h$
cuboid has an odd number of $3\times 3$ planes and is non-polar.
It consists of a closed shell positively charged cluster
plus a compensating electron distributed mainly on corner Na$^+$ ions. }
\end{figure}

We have studied Na$_N$F$_N$ for $N$=2, 4, 6, 9, 18, and 32.
For contrast, we also studied the system Na$_{14}$F$_{13}$ which is a symmetric
$3\times 3 \times 3$ cuboid of $O_h$ symmetry.  Except for $N$=2,
and the ring structure of $N$=4, these are all cuboids, shown in 
Fig. \ref{fig:struc}.  
Vibrational frequencies and infrared activities used to compute
$\alpha_{\rm vib}$ are summarized in Fig.  \ref{fig:vibs}.
Polarizabilities are summarized in Figs. \ref{fig:alpha} and \ref{fig:alpx},
and polarizabilities, formation energies, and permanent dipole moments
are summarized in Table \ref{tab:pol}.
 
\par
\par

\subsection{Na$_2$F$_2$}

This cluster is not a cuboid, but is included because it has previously
been studied experimentally \cite{Eisenstadt,Hartley,Lapshina,Cyvin}
and theoretically \cite{Welch1,Dickey,Modisette}.
Our DFT gives a binding energy relative to two
separated diatomics of 2.4eV.  Previous calculations by Dickey
{\it et al.} \cite{Dickey} gave 2.6eV, which agrees well with 
the experiment of Eisenstadt \cite{Eisenstadt}.
The geometry is a rhombus with Na-F-Na angle 94.7$^{\circ}$
and Na-F bond length 3.96 $a_B$, agreeing with Dickey {et al.}
and with experiments by Hartley and Fink \cite{Hartley} and
Lapshina {\it et al.} \cite{Lapshina}.

The DFT prediction of polarizabilities is given in 
Table \ref{tab:pol}.  The electronic contribution is
surprisingly isotropic.  Per ion pair, the polarizability
$\alpha_{{\rm el},2}$, plotted on Fig. \ref{fig:alpha} is significantly
reduced compared with the diatomic molecule, and closer
to the bulk limit.  The vibrational polarizability has
a large anisotropy dominated by out of plane motions, and
per ion pair, $\bar{\alpha}_{{\rm vib},2}$ is 
much larger than the diatomic and out-of-line with the
results for larger clusters, already close to the bulk value.
The 6 vibrational normal modes consist of three even, Raman
active modes (A$_g$ at 205 cm$^{-1}$, B$_{1g}$ at 326, and
another A$_g$ at 377) and three odd, infrared active modes,
roughly equally intense (B$_{1u}$ at 149 cm$^{-1}$, B$_{3u}$
at 362, and B$_{2u}$ at 373).  In assigning these symmetries,
we used a coordinate system where the $x$-axis passes through
the Na atoms and the $y$ axis through the F atoms.  
These numbers can be read from Fig. \ref{fig:vibs}, which also
shows the relative infrared intensities of the active modes.
Our frequencies agree to better than 10\% with unrestricted Hartree-Fock
calculations \cite{Dickey} and somewhat less well with
shell model calculations \cite{Welch1}.  A matrix-isolation
experiment \cite{Cyvin} has reported two Na$_2$F$_2$
infrared active vibrations above a spectrometer cutoff at 190 cm$^{-1}$.
The experimental numbers 360 and 363 cm$^{-1}$ are close to our values. 
It is the low frequency B$_{1u}$ mode, polarized out of plane, which is
responsible, because of its low frequency,
for the large $zz$-component of vibrational polarizability
in our calculation.

\subsection{Na$_4$F$_4$}

\begin{figure}
\includegraphics[angle=0,width=0.44\textwidth]{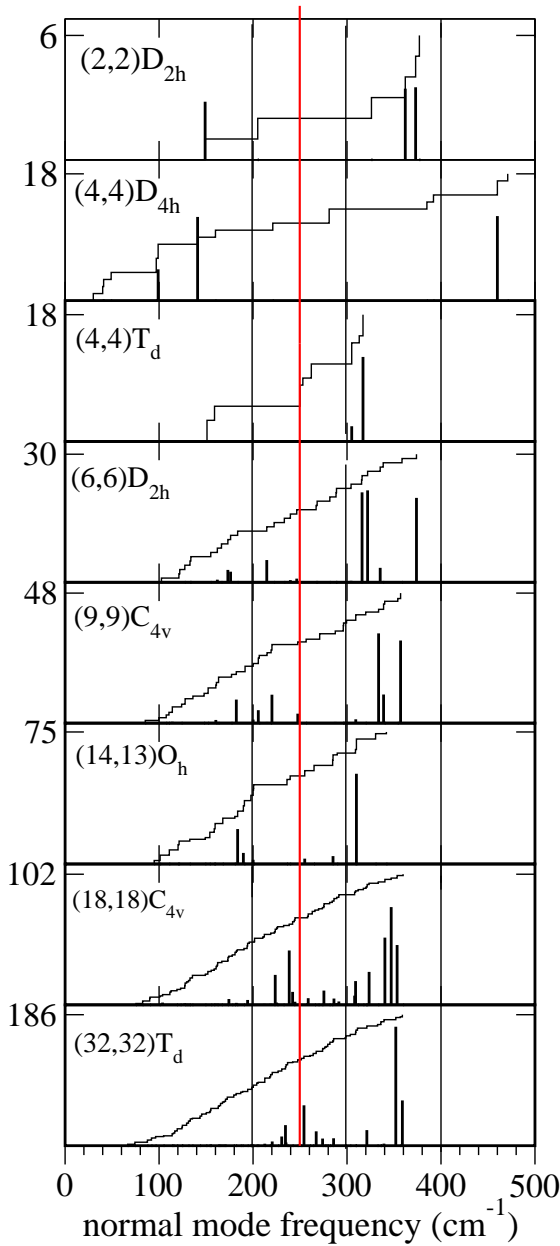}
\caption{\label{fig:vibs} Integrated vibrational densities
of states for various clusters of NaF.  The
degeneracy of each mode is coded in the step increase, as
the spectrum steps upward to the final mode $3N-6$.
Vertical bars are the relative infrared activities of
the various modes; the absolute scale is not given.}
\end{figure}

This cluster has a relatively low-lying structural isomer
with a D$_{4h}$ ring geometry which has been studied theoretically
\cite{Calvo}.  We find the T$_d$ cuboid geometry to be lower
in energy than the ring by $\Delta$=0.75 eV, twice as much as the figure
0.37 eV mentioned by Calvo \cite{Calvo}.  The vibrations of
the ring have a wider range of frequencies
than any of the cuboids, with many low-frequency modes.
We estimate the temperature
of isomerization to be 1000 K in the following way.  If only
two isomers need consideration and if harmonic approximation
is sufficient, and if we ignore rotations, 
then statistical mechanics says that
the fraction of rings $f_R$ should be given by
\begin{equation}
1/f_R= 1+e^{\beta\Delta}\prod_i\frac{\sinh(\beta\hbar\omega_i^R /2)}
{\sinh(\beta\hbar\omega_i^C /2)}
\label{eq:fring}
\end{equation}
where $\omega_i^R$ is the $i$-th vibrational mode of
the ring and $\omega_i^C$ is the $i$-th vibrational mode of
the cuboid.
At low $T=1/\beta k_B$, the ring population is exponentially
small, but as $T$ increases, the lower vibrational frequencies
give an entropic stabilization of the rings.  In the classical
limit $\beta\hbar\omega/2 \ll 1$, the temperature $T_R$ at which
the ring population is 1/2 is
\begin{equation}
k_B T_R = \Delta/\sum_i \log(\omega_i^C/\omega_i^R)
\label{eq:TR}
\end{equation}
Our computed vibrational spectra, shown in
Fig. \ref{fig:vibs}, give the logarithmic
denominator to be 8.5, greatly reducing the temperature of
isomerization.  This reduction factor should be slightly increased
to take into account the higher rotational moments of inertia,
and thus higher rotational entropy, of the rings.

The polarizabilities, shown in Table \ref{tab:pol}, are
surprisingly isotropic for the ring.  Electronic
parts are surprisingly similar for the ring and cube
geometries, while the vibrational polarizability is bigger
by a factor 4 in the ring because of the low frequency
infrared activity shown in Fig. \ref{fig:vibs}.  

\begin{table*}
\caption{\label{tab:pol} Polarizabilities (in \AA$^3$), 
formation energy $E_{\rm f}$
(in eV), and permanent dipole moment $\mu$ (in atomic units $ea_B$)
for NaF molecules and nanocrystals.}
\begin{ruledtabular}
\begin{tabular}{llrrrrrrrrr}
&point&\multicolumn{3}{c}{$\alpha_{\rm el}$}&
\multicolumn{3}{c}{$\alpha_{\rm vib}$}&
\multicolumn{2}{c}{$E_{\rm f}$}&permanent \\
&group&$xx$&$yy$&$zz$&$xx$&$yy$&$zz$&theory&expt.&dipole $\mu$ \\ \hline
NaF & C$_{\infty v}$ & 2.4 & 2.4 & 3.3 & 0 & 0 & 0.9 & 5.2 
& 5.3\footnote{Ref. \onlinecite{Ham}}& 3.07 \\
Na$_2$F$_2$ & D$_{2h}$ & 4.3 & 3.8 & 3.7 & 4.0 & 3.8 & 19.4 &
 6.4 & 6.6 \footnote{Ref. \onlinecite{Eisenstadt}} & 0 \\
Na$_4$F$_4$ & D$_{4h}$ & 7.9 & 7.9 & 7.0 & 36.0 & 36.0 & 41.3 &
 6.6 & & 0 \\
Na$_4$F$_4$ & T$_{d}$ & 7.0 & 7.0 & 7.0 & 10.4 & 10.4 & 10.4 &
 7.0 & & 0 \\
Na$_6$F$_6$ & D$_{2h}$ & 10.0 & 10.0 & 10.6 & 17.8 & 17.7 & 17.5 &
7.2 & & 0 \\
Na$_9$F$_9$ & C$_{4v}$ & 15.1 & 15.1 & 14.2 & 29.1 & 29.1 & 28.7 &
7.3 & & 2.01 \\
Na$_{14}$F$_{13}$ & O$_{h}$ & 313. & 313. & 313. & 24.7 & 24.7 & 24.7 &
 & & 0 \\
Na$_{18}$F$_{18}$ & C$_{4v}$ & 28.2 & 28.2 & 29.8 & 52.8 & 52.8 & 62.7 &
7.5 & & 5.12 \\
Na$_{32}$F$_{32}$ & T$_{d}$ & 49.4 & 49.4 & 49.4 & 89.1 & 89.1 & 89.1 & 
7.6 & & 0 \\
\end{tabular}
\end{ruledtabular}
\end{table*}

\par
\par

An interesting property of the Na$_4$F$_4$ T$_d$
structure is that the ground state has an octupole ($xyz$)
moment, but no quadrupole
moment.  In the point group T$_d$, a vector $(x,y,z$)
and an off-diagonal second-rank symmetric tensor ($yz,zx,xy$)
have the same symmetry.  This means that an applied electric 
field $F_z \hat{z}$ induces both a dipole moment $\mu_z$
and a quadrupole moment $\Theta_{xy}$ \cite{Pullman}.
We calculated the octupole moment, which is equivalent to static
point charges of $\pm 0.866e$ on the ions.  This can be
compared with the static charge $\pm 0.83e$ deduced from the
dipole of the monomer NaF.  The induced quadrupole was
computed to be
\begin{equation}
\Theta_{xy}=\frac{3}{2}\int d^3 r xy\rho(\vec{r})
= (-8.2 a_B^4)F_z.
\label{eq:qin}
\end{equation}

\subsection{Larger Stoichiometric Cuboids}

\begin{figure}
\includegraphics[angle=0,width=0.4\textwidth]{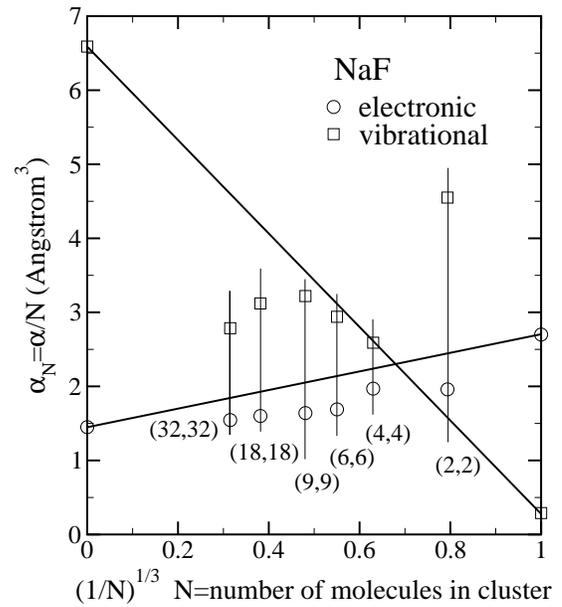}
\caption{\label{fig:alpha} Scaling with $1/N^{1/3}$ 
of the polarizability per NaF molecule $\alpha_N$ of
clusters of various sizes Na$_N$F$_N$.  The diatomic
or monomer ($N=1$) and the dimer ($N=2$)
are not fragments of rocksalt, but the other clusters, $N$=4,6,9, and 18,
are fragments of the bulk crystal ($N=\infty$). 
The straight lines just connect the diatomic to the bulk as if the
corrections were linear in $1/N^{1/3}$.}
\end{figure}

The vibrational eigenfrequencies are shown in Fig. \ref{fig:vibs}.
The polarizabilities are shown in Fig. \ref{fig:alpha},
and contrasted with crystalline compounds in Fig. \ref{fig:alpx}.
We see a non-monotonic but fairly smooth evolution of both vibrational
and electronic polarizabilities with linear dimension of
the cluster, provided we ignore the planar species Na$_2$F$_2$
and D$_{4h}$ Na$_4$F$_4$.  
The Na$_4$F$_4$ and Na$_{34}$F$_{32}$ T$_d$ structures are the 
first two in the series of T$_d$ structures of $2n\times 2n\times 2n$
atoms with isotropic polarizabilities whose properties should
scale smoothly and perhaps rapidly to the bulk limit.  The electronic
polarizability does approach the bulk limit rapidly, but the vibrational
part does not.  There is a subtlety in the vibrational polarizability
of the bulk.  The Lyddane-Sachs-Teller relation shows that the
splitting of TO and LO optical phonon frequencies is tied to the
splitting $\epsilon_{\rm dc}-\epsilon_{\rm opt}$.  In point
group T$_d$ or O$_h$, the infrared optic modes belong to triply
degenerate representations -- there is no distinction between
LO and TO.  The bulk crystal violates this symmetry only
because the distance of spatial variation of the probing electric
field is smaller than the sample size.  Therefore the bulk
limit for $\alpha_{\rm vib}$ may not be reached until
the sample size exceeds the wavelength of the radiation used
to measure $\epsilon_{\rm dc}$. 

We have done three clusters with uniaxial anisotropy.
Na$_6$F$_6$ is first in a sequence of long skinny cuboids
of $2\times 2\times (2n+1)$ structures with D$_{2h}$
symmetry or of $2\times 2\times (2n+2)$ structures with D$_{2d}$
symmetry.  Actually the former group is not strictly uniaxial,
but has weak splitting of the equatorial two-fold axes, which
accounts for a very weak splitting of the $xx$ and $yy$ parts
of the $\alpha$ tensors.  Na$_9$F$_9$ and Na$_{18}$F$_{18}$
are the first two of the C$_{4v}$ sequence of $3\times 3\times 2n$
structures.  The polarizability is relatively isotropic,
except that the vibrational polarizability of Na$_{18}$F$_{18}$
is 20\% enhanced along its long axis.  The clusters with
dimensions $3\times 3\times (2n+1)$ have an odd atom count
and are thus non-stoichiometric.  The O$_h$ $3\times 3\times 3$
cluster Na$_{14}$F$_{13}$ is discussed below; the higher
clusters have D$_{4h}$ symmetry.

\par
\par

\subsection{Permanent Dipole Moments}

The two cuboids of C$_{4v}$ symmetry, Na$_9$F$_9$
and Na$_{18}$F$_{18}$, have permanent dipole moments
just of structural necessity.  Naively, if the cuboid
were an exact sodium chloride structure fragment, with 
the NaF bulk lattice constant of 4.62 \AA, and
the charge of $\pm e$ were assigned to the ions, then
the dipole moments would be 2.31 e\AA \  and 4.62 e\AA \ 
for the smaller and larger cuboids.  However, the 
ions have somewhat distorted positions, and the 
structural dipoles are reduced to 1.34 e\AA \  and
3.16 e\AA \  respectively, assuming unit quantized charges.
The actual DFT values, shown in Table \ref{tab:pol} are
1.06 e\AA \  and 2.71 e\AA, respectively.  These numbers show
that the total dipole is reduced by screening, both 
``vibrational'' and electronic.  Vibrational screening
is larger than electronic, in line with the fact
that vibrational polarizabilities are roughly twice
electronic polarizabilities.  The slab-shaped Na$_9$F$_9$
is more affected by screening than the needle-shaped
Na$_{18}$F$_{18}$, as one expects from
the well-known depolarization effects in slab {\it vs.} needle
shaped samples.  The dipole of Na$_{18}$F$_{18}$ was
previously calculated by Rayane {\it et al.} \cite{Rayane}
to be 7.5 D = 1.84 e\AA, significantly smaller than our
value.  Their calculation was part of an investigation of
non-stoichiometric Na$_N$F$_{N-1}$ species such as Na$_{18}$F$_{17}$.
Only the outer electron was treated by quantum mechanics.  The
stoichiometric species has no outer electron, so apparently their
number is based on a classical polarizable ion model, which
apparently overscreens.

\subsection{Non-stoichiometric O$_h$ Na$_{14}$F$_{13}$}

This is such an interesting cuboid that we could not
resist departing from our main theme of polarizability
evolution in stoichiometric cuboids.  
The last electron resides outside the closed shell
of Na$_{14}^+$F$_{13}^-$.  Alkali halides with one electron
outside a closed shell have attracted attention 
\cite{Rayane,Fatemi} since the
prediction by Landman {\it et al.} that
such clusters might have a second-order Jahn-Teller instability.
This, and the subsequent studies by Durand {\it et al.} 
\cite{Durand1,Durand2}
kept only a single electron in the quantum part of the dynamics,
while Ochsenfeld and Ahlrichs \cite {Ochsenfeld}
have studied several alkali halides using all-electron theory.

The single outer electron gives rise to a giant
electronic polarizability shown in Table \ref{tab:pol}.
On a per molecule basis, the total polarizability of 313
\AA$^3$, shared by 13.5 molecules, is larger by 15 than 
the interpolated value $\sim 1.5$ seen in Fig. \ref{fig:alpha}.
On an absolute basis, it is larger by 13.5 than the polarizability
of the neutral Na atom.  On a per molecule basis, the vibrational
polarizability of 24.7 is smaller by 0.6 than the interpolated
value of the stoichiometric clusters.  Apparently the loose last
electron is exceedingly polarizable, and partly screens the
change in dipole per unit displacement that causes vibrational
polarizability according to Eq. (\ref{eq:avib}).

The large polarizability gives one a right to expect
that Na$_{14}$F$_{13}$ might distort 
to a lower symmetry polar structure such as C$_{3v}$,  
as was discovered theoretically in a restricted model
by Landman {\it et al.} \cite{Landman} in the Na$_{14}$Cl$_{13}$ cuboid.
In a naive view, point-polarizable ions have a density-dependent
critical $\alpha_{\rm el,c}$ beyond which ions
polarize spontaneously.  In cubic crystals it is $3/4\pi n$
which equals 5.88 \AA$^3$ for the density $n$ in bulk NaF.
In fragments $\alpha_c$ approaches the bulk limit rapidly \cite{Allen}.
The per molecule electronic polarizability of Na$_{14}$F$_{13}$
is way beyond that limit, and thus unstable in this model.  
However, the giant polarizability is almost all associated with the
delocalized last electron, making the Clausius-Mossotti analysis
inappropriate.  

The highest occupied molecular orbital 
(HOMO) is non-degenerate
but lies not far below other levels with which it is forbidden
by symmetry to mix.  The second-order Jahn-Teller effect is 
a distortion motivated by the ``desire'' for
such a mixture in order to lower (quadratically in distortion amplitude)
the HOMO energy.  Kristoffel and Konsin \cite{Kristoffel} argue that
such a mechanism is at work in the ferroelectric distortion of
GeTe crystals, involving the mixing between states
just below and just above the narrow gap of this IV-VI semiconductor
with two more electrons than are needed to fill the usual $s-p$ shell.
The instability of Landman {\it et al.} \cite{Landman} 
is a nanocrystalline analog.

We find that this instability {\bf does not happen} in Na$_{14}$F$_{13}$.
This contradicts other theories and the interpretation 
by Rayane {\it et al.} \cite{Rayane} of their experiment.
Theories which keep only the extra outer electron
are not reliable enough if the energy balance is delicate.
Rayane {\it et al.} \cite{Rayane} found a very large deflection of 
neutral Na$_{14}$F$_{13}$ particles in electric field gradients. 
There was a moderate $T$-dependence, but not the full $1/T$
of the usual Debye-Langevin polarizability $\mu^2/3k_B T$ of a molecule with a
permanent dipole $\mu$.  Nevertheless, the authors believe their data
to be consistent with a permanent dipole, although they
do not claim that their analysis is firm.  
Our finding is that the electronic polarizability is indeed
extremely large and will give large deflections, but there is no 
symmetry-breaking or permanent dipole moment.  We believe the data
may be consistent with this alternate interpretation.

\par
\begin{figure}[top]
\includegraphics[angle=0,width=0.4\textwidth]{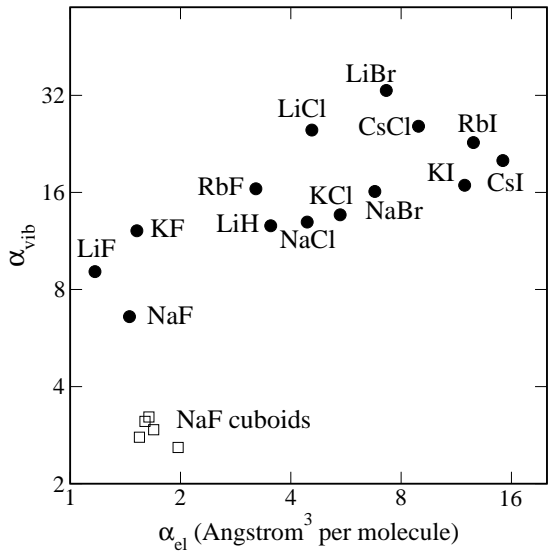}
\caption{\label{fig:alpx} Log-log plot of vibrational polarizability (vertical)
{\it vs.} electronic polarizability (horizontal).  Open squares
are the NaF cuboid clusters.  Closed circles are from crystal data.}
\end{figure}
\par

\section{discussion}

\subsection{Formation energy}
We define the formation energy $E_f$ as the energy per ion pair
relative to the energy of separated neutral atoms.  
Values of $E_f$ are given in Table \ref{tab:pol}.  The increasing
formation energy with cluster size can be organized phenomenologically
according to the numbers of ions with various geometric sites.  We
distinguish corner, edge, face, and interior ions.  In the Na$_N$F$_N$
cuboids these sites occur as pairs of oppositely charged ions.  
For example, Na$_4$F$_4$ has 4 corner pairs, Na$_6$F$_6$ has
4 corners and 2 edges, Na$_9$F$_9$ has 4 corners, 4 edges, and one face,
and Na$_{18}$F$_{18}$ has 4 corners, 8 edges, 5 faces, and 1 interior
ion pair.  The values of $E_f$ then correspond to the assignment
7.01 eV for corner pairs, 7.53 eV for edge pairs, 7.63 eV for face
pairs, and 9.11 eV for interior pairs.  This last is close to the
experimental bulk value of 9.30 eV.  These assignments then predict
$E_f$ for larger cuboids.  Na$_{32}$F$_{32}$ has 4 corners, 
12 edges, 12 faces, and 4 interior pairs, so is predicted to have
$E_f$=7.70 eV, very close to the computed DFT value of 7.6 eV.

\subsection{Trends of Polarizability}

The trend to smaller electronic polarizability as size increases
should certainly be attributed to enhanced stabilization of the
filled $p$ shell as the ion acquires larger numbers of neighboring
counterions.  The HOMO and LUMO states of a finite cluster are
located mostly at the surface.  
The HOMO-LUMO gap is reduced in clusters relative to bulk,
corresponding to a cluster analog of surface localized states
on a bulk surface.  As the cluster size increases, electronic
spectral weight gets pushed upward toward the bulk band
gap of 11eV \cite{Rao}.
In bulk alkali halides, an excess electron is a delocalized
carrier, but an excess hole self-traps on a dimerized
anion pair (V$_K$ center).  On clusters, one would
expect the trapped hole to lie on surface anions.
The excess electron, after migrating to the surface, might also
self-trap.  This would be a ``pseudo-Jahn-Teller polaron''.  The
fact that the second-order Jahn-Teller effect does not occur
in Na$_{18}$F$_{17}$ (according to us) suggests that the
surface electron might not trap but remain delocalized.
We note that localization of the excess electron could be enhanced
within a formalism where the electronic self-interaction (SI) is fully 
corrected \cite{Perdew}.  Further investigation of this effect within an
SI-free framework would be necessary to rule out the 
possibility of surface trapping.

The enhanced vibrational polarization of larger clusters is
harder to explain.  Large electronic polarizability should
diminish vibrational polarizability by screeing.  The dynamical
effective charge matrix (change of electrical dipole per unit
displacement) is reduced by the polarizability of the ions which
are displacing (see Eq. \ref{eq:diat}.)  We find little support
for this idea in the trends of dielectric constants of other
alkali halides, shown in Fig. \ref{fig:alpx}.  Here the value of
$\alpha$ per molecule is $(\epsilon-1)/4\pi n$, with the optical
dielectric constant giving the $\alpha_{\rm el}$ as before.
The vibrational polarizability seems 
roughly independent of the electronic polarizability.

\begin{acknowledgements}
We thank Y.-R. Chen and D. O. Welch for helpful discussions,
and Y.-R. Chen for help with the manuscript.  Work at
Stony Brook was supported in part by NSF Grant DMR0089492.  C. S. was 
supported by a student fellowship from the German government.
T. B. and M. P. were supported in part by the ONR and the
HPCMO and CHSSI initiatives.
\end{acknowledgements}

\end{document}